\documentstyle[12pt,psfig]{article}
\catcode`\@=11
\def\marginnote#1{}
\newcount\hour
\newcount\minute
\newtoks\amorpm
\hour=\time\divide\hour by60
\minute=\time{\multiply\hour by60 \global\advance\minute by-
\hour}
\edef\standardtime{{\ifnum\hour<12 \global\amorpm={am}%
    \else\global\amorpm={pm}\advance\hour by-12 \fi
    \ifnum\hour=0 \hour=12 \fi
    \number\hour:\ifnum\minute<100\fi\number\minute\the\amorpm}}
\edef\militarytime{\number\hour:\ifnum\minute<100\fi\number\minute}
\def\draftlabel#1{{\@bsphack\if@filesw {\let\thepage\relax
  \xdef\@gtempa{\write\@auxout{\string
    \newlabel{#1}{{\@currentlabel}{\thepage}}}}}\@gtempa
    \if@nobreak \ifvmode\nobreak\fi\fi\fi\@esphack}
     \gdef\@eqnlabel{#1}}
\def\@eqnlabel{}
\def\@vacuum{}
\def\draftmarginnote#1{\marginpar{\raggedright\scriptsize\tt#1}}
\def\draft{\oddsidemargin -.5truein
        \def\@oddfoot{\sl preliminary draft \hfil
        \rm\thepage\hfil\sl\today\quad\militarytime}
        \let\@evenfoot\@oddfoot \overfullrule 3pt
        \let\label=\draftlabel
        \let\marginnote=\draftmarginnote

\def\@eqnnum{(\theequation)\rlap{\kern\marginparsep\tt\@eqnlabel}%
\global\let\@eqnlabel\@vacuum}  }
\def\preprint{\twocolumn\sloppy\flushbottom\parindent 1em
        \leftmargini 2em\leftmarginv .5em\leftmarginvi .5em
        \oddsidemargin -.5in    \evensidemargin -.5in
        \columnsep 15mm \footheight 0pt
        \textwidth 250mmin      \topmargin  -.4in
        \headheight 12pt \topskip .4in
        \textheight 175mm
        \footskip 0pt

\def\@oddhead{\thepage\hfil\addtocounter{page}{1}\thepage}
        \let\@evenhead\@oddhead \def\@oddfoot{} \def\@evenfoot{}
}
\def\titlepage{\@restonecolfalse\if@twocolumn\@restonecoltrue\onecolumn
     \else \newpage \fi \thispagestyle{empty}\c@page\z@
        \def\thefootnote{\fnsymbol{footnote}} }
\def\endtitlepage{\if@restonecol\twocolumn \else  \fi
        \def\thefootnote{\arabic{footnote}}
        \setcounter{footnote}{0}}  
\catcode`@=12
\relax
\def\be{\begin{equation}}
\def\ee{\end{equation}}
\def\bea{\begin{eqnarray}}
\def\eea{\end{eqnarray}}

\def\NPB#1#2#3{{\it Nucl.~Phys.} {\bf{B#1}} (19#2) #3}
\def\PLB#1#2#3{{\it Phys.~Lett.} {\bf{B#1}} (19#2) #3}
\def\PRD#1#2#3{{\it Phys.~Rev.} {\bf{D#1}} (19#2) #3}
\def\PRL#1#2#3{{\it Phys.~Rev.~Lett.} {\bf{#1}} (19#2) #3}

\newcommand{\beq}{\vspace{2mm}\begin{eqnarray}}
\newcommand{\eeq}{\end{eqnarray}\vspace{2mm}}

\relax

\begin{document}
\textheight 23cm
\textwidth 15.5cm
\topmargin-2.5cm

\begin{flushright}
CERN-TH/98-21\\
hep--ph/9801376 \\
\end{flushright}
\vskip 1. in

\begin{center}
{\Large\bf On the Soft Supersymmetry Breaking
Parameters in Gauge-Mediated Models}\\

\vskip .3 in
{\bf C.E.M. Wagner} 
\vskip .5 in
CERN, TH Division, CH--1211 Geneva 23, Switzerland
\vskip .1 in
\vskip 2.cm
{\bf Abstract}
\end{center}
\begin{quote}
Gauge mediation of supersymmetry breaking in the observable
sector is an attractive idea, which naturally alleviates the
flavour changing neutral current problem of supersymmetric
theories. Quite generally, however, the number and quantum
number of the messengers are not known; nor is their characteristic
mass scale determined by the theory. Using the recently proposed
method to extract supersymmetry-breaking parameters from wave-function 
renormalization, we derived general formulae
for the 
soft supersymmetry-breaking parameters in the observable sector,
valid in the small and moderate $\tan\beta$ regimes, for the 
case of split messengers.
The full leading-order effects of  top Yukawa and gauge couplings  
on the soft supersymmetry-breaking parameters are included. We
give a simple interpretation of the general formulae
in terms of the renormalization group evolution of the soft
supersymmetry-breaking parameters. As a by-product of this analysis,
the one-loop renormalization group evolution of the
soft supersymmetry breaking parameters is obtained for
arbitrary boundary conditions of the scalar and gaugino mass
parameters at high energies.
\end{quote}
\vskip.8cm
\begin{flushleft}
CERN-TH/98-21 \\
January 1998
\end{flushleft}

\setcounter{footnote}{0}
\setcounter{page}{0}
\newpage

\noindent
\textheight 23.5 cm
\textwidth 15 cm
\voffset -1 cm
\hoffset -0.5 cm
\topmargin -0.0 cm

\section{Introduction}
Low energy supersymmetric theories have attracted much attention
in the past years, since they can solve the hierarchy
problem of the Standard Model and
provide an effective low-energy
description of grand unified theories,
describing all known interactions, including gravity~\cite{NILLES}. 
In spite
of these remarkable properties, since
supersymmetry is broken at low energies, a complete 
phenomenological description of the low energy physical phenomena
may only be obtained after the exact mechanism
leading to the breakdown of supersymmetry in the observable
sector is known. It was long ago realized that more
important than  the nature of this mechanism is the
way the supersymmetry-breaking effects are transmited
to the observable sector. 

One of the most interesting ideas concerning the origin of
supersymmetry breaking is that it is  transmitted to the
observable sector through superfields which couple to the 
Standard Model superfields only via 
gauge interactions~\cite{DINE},\cite{Review}. 
These
superfields, usually called messengers, are coupled to the sector
in which supersymmetry is originally broken (in a spontaneous
or dynamical way). The supersymmetry-breaking effects in the
observable sector are obtained by decoupling the messenger states,
thus generating an effective theory at low energies, which contain
the information about the soft supersymmetry-breaking parameters.
In the simplest cases, the coupling of the supersymmetry-breaking 
sector with the messengers may be parametrized by the
introduction of auxiliary chiral superfields $X_I$, coupled to the
messengers $\Phi_I$, and with non-vanishing expectation 
values in their scalar ($X_I$) and auxiliary ($F_{X_I}$) components~\cite{DINE}
\footnote{In the following, we shall assume the absence of off
diagonal $F$-terms in the messenger flavor space~\cite{DG}.}. 
For simplicity of
notation, in the following, we shall omit the brackets in the
definition of the field vacuum expectation values.

One of the phenomenological properties of gauge-mediated 
supersymmetry-breaking models
is that fields with the same quantum numbers under the Standard
Model gauge group acquire the same soft supersymmetry breaking
masses, leading to a natural suppression of the flavour changing
neutral current effects via a super-GIM mechanism. Moreover,
in order to preserve the successful unification of couplings in the
minimal supersymmetric standard model (MSSM), the messengers are
assumed to form complete representations of the group $SU(5)$ and,
for these fields to acquire explicit SUSY conserving masses, 
that they come
in vector representations ($5 \; \bar{5}$, $10 \; \bar{10}$, etc.). 
In addition, it is usually assumed that the
dynamics leading to the non-vanishing $X_I$'s and $F_{X_I}$'s preserves
the $SU(5)$ symmetry, and that all messengers acquire a single
mass scale, modulo wave-function renormalization 
and supersymmetry breaking effects. Although
these assumptions lead to phenomenologically acceptable models, 
the way supersymmetry-breaking effects are transmitted to the
messenger sector may violate the
exact $SU(5)$ relations \cite{Martin},\cite{Stuart},\cite{AG}. 
Moreover, the messengers
may acquire different masses, provided that, apart from the usual
doublet-triplet splitting in the Higgs sector,  
the hierarchy of masses
between different SU(5) multiplet states is not extremely large,
in order to preserve the unification of couplings (other possibilities
are discussed in Ref.~\cite{Martin}).

An example of these generalized scenarios was provided in Ref.~\cite{Stuart},
where the messengers that transmit supersymmetry breaking proceed
from two 10 $\bar{10}$ multiplets of $SO(10)$, one of which contains the 
usual Higgs doublets. In this case, the triplets and the heavy
doublets act as messengers, which acquire different masses and
feel the supersymmetry breaking effects in different ways. The
phenomenological consequences of this kind of scenarios may be
very different from the minimal one. In the particular case
of Ref.~\cite{Stuart}, light gluinos appear in the spectrum, due to
a suppression of the supersymmetry-breaking effects in the colour
sector, originated from
the Higgs triplet states. Since no such 
suppression factor appears in the terms originated from the heavy doublet
states,  particles with electroweak
quantum numbers acquire much larger masses than the coloured ones. 

Our purpose in this letter is not to concentrate on any particular model,
but to present general formulae for an arbitrary messenger
sector, independently of their quantum numbers. In section 2, we review
the method to extract the supersymmetry-breaking terms from wave-function 
renormalization effects, presenting a generalization of
the method derived in Ref.~\cite{RG} for the case of many messengers.
In section 3 we present an example, with a messenger sector consisting
in a $5$ $\bar{5}$ of SU(5) and with doublet and triplet components 
whose masses are split in an arbitrary way. In section 4 we present
the general formulae. We reserve section 5 for our conclusions. Relevant
formulae for the derivation of the expressions presented in sections
3 and 4 are given in the appendix.

\section{Method to extract supersymmetry-breaking terms}

Let us start with the effective superpotential
\begin{equation}
W = \sum_i X_I \bar{\Phi}_I \Phi_I ,
\end{equation}
where, as mentioned in the introduction, we shall assume that
both $X_I \neq 0$ and $F_{X_I} \neq 0$. The fermion
components of the messengers  $\Phi_I$ acquire a mass equal
to $X_I$ in the background of the $X_I$ field, while the
scalar components acquire supersymmetry breaking masses
$m_{S}^2 = X_I^2 \pm F_{X_I}$.   
Observe that the sign of the
vacuum expectation values is not fixed a priori, what, of
course, has no relevance at this stage. 

Under these conditions, the values of the soft supersymmetry-breaking 
parameters in the observable sector
may be obtained by a simple generalization
of the method developed in Ref.~\cite{RG}. Let us briefly discuss
this method. The gauge superfield kinetic term may be parametrized
by 
\begin{equation}
{\cal{L}}_W = \int d^2\theta S_j W_{\alpha}^j W^{\alpha}_j ,
\end{equation}
where Re$(\langle S_j \rangle) = \alpha_j^{-1}/4\pi$, 
while $-\langle F_{S_j}/(2 S_j)\rangle$ 
is the renormalized gaugino-mass parameter, with $j$ a group 
index parameter. Throughout this paper, summation over repeated
indices will be implicitly understood.  
Owing to the
presence of the messengers, the vacuum expectation values 
of $S_j$ and $F_{S_j}$ are
a function of $X_I$ and $F_{X_I}\equiv F_I$. Assuming that all
$F_I \ll X_I^2$, and 
at the first order in $F_I/X_I^2$ ,
one can easily show that
\begin{equation}
M_j = -\frac{1}{2}\frac{\partial \ln S_j}{\partial \ln X_I} \frac{F_I}{X_I} .
\end{equation}
In general, the sign and phase of the different contributions 
to the gaugino masses may
be important, since cancellations between 
them may occur and, in addition, 
non-vanishing CP-violating phases may be induced in the
low-energy theory.

For the scalar masses, the kinetic terms arise from
\begin{equation}
{\cal{L}} = \int d^4\theta Z_Q(X_J) \Phi_Q \bar{\Phi}_Q ,
\end{equation}
where $Q$ denotes any arbitrary superfield in the observable sector.
At the lowest order in $F_I/X_I^2$,
the renormalized soft supersymmetry-breaking
scalar masses are now given by
\begin{equation}
m^2_Q = - \frac{\partial^2 \ln Z_Q(X_I)}{\partial X_J \partial X_I^{\dagger}}
\frac{F_J F_I^{\dagger}}{X_J X_I^{\dagger}} .
\label{scalarm}
\end{equation}
Observe that the $X_I$ dependence of $Z_Q$ comes through the
gauge-coupling dependence of the wave-function renormalization. 

Finally, let us consider the trilinear and bilinear terms in the
scalar potential, defined through
\begin{equation}
V = A_Q \Phi_Q \frac{\partial W}{\partial \Phi_Q} ,
\end{equation}
where $A_Q$ denotes the soft supersymmetry-breaking
parameters and, as before, 
we have used the same notation for the superfield as for the
scalar terms. The expression for $A_Q$ is given by
\begin{equation}
A_Q = \frac{\partial \ln Z_Q}{\partial \ln X_I} \frac{F_I}{X_I}.
\end{equation}

It is interesting to notice that,
in all  cases considered above,
the supertrace Str${\cal M}^2$ vanishes 
in the messenger sector~\cite{PT}.
The treatment of the case with non-vanishing Str${\cal M}^2$ in
the messenger sector demands the inclusion of the next-to-leading
effects in the wave-function renormalization, an extension
that is beyond the scope of this article. 

\section{General Formulae for the Soft Terms. \\
An Example}

Let us start with an example, which has all the necessary 
features to derive the general formulae for the soft terms
in gauge-mediated models. Let us imagine that the messengers
belong to one set of $5$ $\bar{5}$ representations of $SU(5)$, 
but that
the doublets and triplets have different masses $X_I$, and,
in general, different $F_I$ terms. The theory then has three
different energy regimes: above the largest of the two
messenger masses, the theory has the MSSM plus the $5$ $\bar{5}$
with the corresponding $\beta_i$ function coefficients for the
couplings $\alpha_i$, which we shall denote by $b_i$. For 
definiteness, we shall take an example in which $X_2 > X_3$.
At scales below $X_2$, but above $X_3$,
the SU(2) messenger doublets decouple and we are left with a 
theory with the MSSM particle content plus two $SU(3)$ colour
triplets with the quantum numbers of the right 
handed down quark and its charge conjugate.
The corresponding $\beta_i$ function coefficients will be
denoted by $b_i^{3\bar{3}}$. 
Below $X_3$, we are left with the MSSM, with $\beta_i$
function coefficients equal to $b_i^{MSSM}$.
In this particular example, the numerical values of these
coefficients are given by the following table
\begin{center}
\begin{tabular}{|c | c | c | c|} \hline
& Full theory &  MSSM + $3$ $\bar{3}$ &  MSSM \\
\hline\hline
$b_3$ &   $-2$   &     $-2$     &      $-3$ \\

\hline
$b_2$ &   2   &     1      &    1       \\
\hline
$b_1$  &   38/5   &     7      &    33/5      \\
\hline
\end{tabular}
\end{center}

More important than the particular values of the $\beta_i$ functions
in the three regimes are the  contributions of the messenger
states to the $\beta_i$ function coefficients, which can be obtined
by the information given in the above table. We shall call $B_2^i = 3/5,1,0$
the $SU(2)_L$ doublet contribution to these
coefficients, and $B_3^i = 2/5,0,1$ the $SU(3)$ triplet contribution
for $i = 1,2,3$, respectively.

Let us begin with the gaugino masses. The gauge couplings dependence on
the messenger scales may be simply obtained by renormalization group
methods. At the one-loop level, we get
\begin{eqnarray}
\frac{1}{\alpha_j(X_2)} = \frac{1}{\alpha_j(M_G)} + \frac{b_j}{4\pi}
\ln\left(\frac{M_G^2}{|X_2|^2}\right) ,
\nonumber\\
\frac{1}{\alpha_j(X_3)} = \frac{1}{\alpha_j(X_2)} + \frac{b_j^{3\bar{3}}}{4\pi}
\ln\left(\frac{|X_2|^2}{|X_3|^2}\right) ,
\nonumber\\
\frac{1}{\alpha_j(\mu)} = \frac{1}{\alpha_j(X_3)} + \frac{b_j^{MSSM}}{4\pi}
\ln\left(\frac{|X_3|^2}{|\mu|^2}\right) .
\end{eqnarray}
Applying the above rules, and using the holomorphicity of
$S_j(X_I,\mu)$, the gaugino masses $M_j$ are given by
\begin{equation}
M_j(\mu) = \frac{B_2^j \alpha_j(\mu)}{4\pi}  \frac{F_2}{X_2} + 
           \frac{B_3^j\alpha_j(\mu)}{4\pi}  \frac{F_3}{X_3}.
\end{equation}
Numerically, we hence obtain,
\begin{equation}
M_3(\mu) = \frac{\alpha_3(\mu)}{4\pi} \frac{F_3}{X_3}, \;\;\;\;\;\;\;\;\;
M_2(\mu) = \frac{\alpha_2(\mu)}{4\pi} \frac{F_2}{X_2}, \;\;\;\;\;\;\;\;\;
M_1(\mu) = \frac{\alpha_1(\mu)}{4\pi} 
\left( \frac{3 F_2}{5 X_2} + \frac{2 F_3}{5 X_3}
\right) ,
\end{equation}
a result that agrees with the one obtained, for the same conditions, in
Ref. \cite{Martin}. Observe that the gaugino masses receive contributions
from the decoupling of both messenger states, which are proportional
to  the beta function contribution of the messenger $X_I$ and to
$\alpha_i(X_I)$, while their dependence
on $\alpha_i(\mu)$ is  a reflection of the evolution of the gaugino
masses at the one-loop level
\begin{equation}
\frac{d M_i}{dt} = \frac{b_i M_i \alpha_i}{4\pi},
\;\;\;\;\;\;\;\;\;\; \;\;\;\;\;\;
\frac{d}{dt} \left(\frac{M_i}{\alpha_i}\right) = 0,
\label{gauginomev}
\end{equation} 
where $t = \ln \mu^2$.
For future purposes, it is useful to define the gaugino masses at
the scales $X_2$ and $X_3$, which are given by
\begin{eqnarray}
M_j(X_2) &=& \frac{B_2^j \alpha_j(X_2)}{4\pi} \frac{F_2}{X_2},
\nonumber\\
M_j(X_3) & = & \frac{B_2^j \alpha_j(X_3)}{4\pi}  \frac{F_2}{X_2} + 
           \frac{B_3^j\alpha_j(X_3)}{4\pi}  \frac{F_3}{X_3},
\label{gaugino23}
\end{eqnarray}
respectively.
The gaugino masses at the scale $X_2$ come  from the 
decoupling effects of the doublet messengers, while $M_j(X_3)$
comes in part from the running between $X_2$ and $X_3$ 
(first term in the second line of Eq. (\ref{gaugino23}))
and in part
from the decoupling effects of the colour triplet messengers
(last term in the second line of Eq. (\ref{gaugino23})).

The scalar sector is more interesting. If all Yukawas are neglected,
the wave-function renormalization  is given by
\begin{equation}
\ln \left(\frac{Z_Q(X_I,X_I^{\dagger},\mu)}{Z_Q(M_G)}\right) =
-\frac{2 c_Q^i}{b_i} \ln\left(\frac{\alpha_i(M_G)}{\alpha_i(X_2)}\right) -
\frac{2 c_Q^i}{b_i^{3\bar{3}}} 
\ln\left(\frac{\alpha_i(X_2)}{\alpha_i(X_3)}\right) -
\frac{2 c_Q^i}{b_i^{MSSM}} \ln\left(\frac{\alpha_i(\mu)}{\alpha_i(X_3)}\right),
\end{equation}
where $c_Q^i$ is the quadratic Casimir of the superfield 
$Q$ under the $i$ gauge group, which, for a fundamental representation
of $SU(N)$ takes the value $c_Q = (N^2-1)/2N$, while $c_Q = Y^2/4$ for
$U(1)$. Observe that we are implicitly working with a normalization
of the gauge couplings
consistent with their unification  at high energies, so
$\alpha_1 = 5/3 \; \alpha_1^{SM}$, and $Y^2/4 \equiv 3/5 \times 
(Q - T_3)^2$.
Applying the above rules, one obtains
\begin{eqnarray}
-\frac{\partial^2 \ln Z_Q}{\partial\ln X_2 \partial \ln X_2^{\dagger}} &=&
- \frac{2 c_Q^i}{16\pi^2} \left[-B_2^i \alpha_i^2(X_2) 
+ \frac{(B_2^i)^2}{b_i^{3\bar{3}}}\left(\alpha_i^2(X_3) - \alpha_i^2(X_2)
\right) 
\right.
\nonumber\\
& + & 
\left.
\frac{(B_2^i)^2}{b_i^{MSSM}}\left(\alpha_i^2(\mu) - \alpha_i^2(X_3)
\right) \right] ,
\nonumber\\
-\frac{\partial^2 \ln Z_Q}{\partial\ln X_2 \partial \ln X_3^{\dagger}} &=&
- \frac{2 c_Q^i}{16\pi^2} \left[
\frac{B_2^i B_3^i}{b_i^{MSSM}}\left(\alpha_i^2(\mu) - \alpha_i^2(X_3)
\right) \right] ,
\nonumber\\
-\frac{\partial^2 \ln Z_Q}{\partial\ln X_3 \partial \ln X_3^{\dagger}} &=&
- \frac{2 c_Q^i}{16\pi^2} \left[-B_3^i \alpha_i^2(X_3) 
+ \frac{(B_3^i)^2}{b_i^{MSSM}}\left(\alpha_i^2(\mu) - \alpha_i^2(X_3)
\right) \right] .
\label{scalarev}
\end{eqnarray}
Taking into account Eqs. (\ref{scalarm}) and (\ref{gaugino23}),
the scalar
masses at the scale $\mu$,
can be defined 
\begin{eqnarray}
m_Q^2(\mu) & = & \frac{2 c_Q^i}{16 \pi^2} B_2^i \alpha_i^2(X_2)
\left(\frac{F_2}{X_2}\right)^2
- 2 c_Q^i \frac{M_i^2(X_2)}{\alpha_i^2(X_2)} 
\left[\frac{\alpha_i^2(X_3) - \alpha_i^2(X_2)}{b_i^{3\bar{3}}}
\right]
\nonumber\\
& + & \frac{2 c_Q^i}{16 \pi^2} B_3^i \alpha_i^2(X_3)
\left(\frac{F_3}{X_3}\right)^2
-2 c_Q^i \frac{M_i^2(X_3)}{\alpha_i^2(X_3)} 
\left[\frac{\alpha_i^2(\mu) - \alpha_i^2(X_3)}{b_i^{MSSM}}
\right] .
\label{mq2}
\end{eqnarray}
The above expression has a simple interpretation. 
The first terms 
in the first and second lines of 
Eq. (\ref{mq2}) are  
the threshold contributions of the messenger
of mass $X_2$ and $X_3$ 
to the scalar soft supersymmetry-breaking mass parameters, respectively.
The second terms in these lines
are  the terms resulting from the evolution
of the scalar mass parameters from $X_2$ to $X_3$ and from
$X_3$ to $\mu$, respectively. 
Indeed, the renormalization group
equations for the scalar mass parameters are  given by
\begin{equation}
\frac{d m_Q^2}{dt} = -4 c_Q^i \frac{\alpha_i M_i^2}{4\pi} .
\label{scalarmev}
\end{equation}
Using this equation, plus the evolution of the gaugino mass
parameters, Eq. (\ref{gauginomev}), it is easy to derive the last terms in the
first and second lines of Eq. (\ref{mq2}).

The scalar masses at the
scale $X_2$, $X_3$ are given by
\begin{eqnarray}
m_Q^2(X_2) & = & \frac{2 c_Q^i}{16 \pi^2} B_2^i \alpha_i^2(X_2)
\left( \frac{F_2}{X_2} \right)^2
\nonumber\\
m_Q^2(X_3) & = & m_Q^2(X_2) 
- 2 c_Q^i \frac{M_i^2(X_2)}{\alpha_i^2(X_2)} 
\left[\frac{\alpha_i^2(X_3) - \alpha_i^2(X_2)}{b_i^{3\bar{3}}}
\right]
\nonumber\\
& + & \frac{2 c_Q^i}{16 \pi^2} B_3^i \alpha_i^2(X_3)
\left( \frac{F_3}{X_3} \right)^2 .
\label{mq2xi}
\end{eqnarray}

\subsection{Yukawa Coupling Effects}

The most interesting contributions to the evolution of the scalar 
masses come from the Yukawa terms. The top quark Yukawa effects
in the renormalization group evolution of the Higgs scalar
masses is indeed essential to induce the radiative breaking of
the electroweak symmetry \cite{IL}--\cite{CCOPW}. 
The top Yukawa effects on the 
supersymmetry-breaking parameters
may be obtained by
studying the evolution of the wave-function renormalization 
\begin{equation}
\frac{d \ln Z_Q}{dt} = \frac{ 2 c_Q^i \alpha_i}{4\pi} -
\frac{d_Q^t \alpha_t}{4\pi} ,
\label{wave}
\end{equation}
where $\alpha_t = h_t^2/4\pi$,  $d_{Q_L}^t = 1$, $d_{U_R}^t = 2$,
and $d_{H_2}^t = 3$, $Q_L$ and $U_R$  are the third
generation left-handed quark doublet and right-handed up quark,
respectively, and $H_2$ is
the Higgs that couples to up-type quarks superfields 
at tree level. In the above, we have ignored the bottom Yukawa coupling,
considering it to be small compared with the top Yukawa coupling. 
The bottom Yukawa coupling effects on the evolution of the soft
supersymmetry breaking parameters may be quite relevant if 
$\tan\beta$ is large, $\tan\beta \simeq m_t/m_b$ \cite{RHS}--\cite{RS}.
Hence,
our analysis is valid in the small and moderate $\tan\beta$ regimes.
To obtain the form of the wave-function renormalization,
Eq. (\ref{wave})  must be complemented with
the evolution of the Yukawa coupling
\begin{equation}
\frac{d \alpha_t}{dt} = -\frac{\alpha_t}{4\pi}\left[c_t^i \alpha_i - 
\left(d_{Q_L}^t + d_{U_R}^t + d_{H_2}^t\right) \alpha_t \right] ,
\end{equation}
where $c_t^i = 2 (c_{Q_L}^i + c_{U_R}^i + c_{H_2}^i)$, while
$\sum_{Q=U_R,Q_L,H_2} d_{Q}^t = 6$.  

The top-quark Yukawa coupling at the different scales
is given by
\begin{eqnarray}
\alpha_t(X_2) & = & \frac{\alpha_t(M_G) E(X_2)}{1 - \frac{6 \alpha_t(M_G)}
{4\pi} F(X_2)},
\nonumber\\
\alpha_t(X_3) & = & \frac{\alpha_t(X_2) E'(X_3)}{1 - \frac{6 \alpha_t(X_2)}
{4\pi} F'(X_3)},
\nonumber\\
\alpha_t(\mu) & = & \frac{\alpha_t(X_3) E''(\mu)}{1 - \frac{6 \alpha_t(X_3)}
{4\pi} F''(\mu)},
\end{eqnarray}
where $dF/\ln t = E$, $dF'/dt'= E'$, $dF''/dt'' = E''$, with
$X_2 \leq t \leq M_G$, $X_3 \leq t' \leq X_2$, $\mu \leq t'' \leq X_3$, 
respectively. The functions $E$, $E'$ and $E''$ are defined by
\begin{equation}
E = \prod_i \left[ \frac{\alpha_i(t)}{\alpha_i(M_G)}\right]^{-c_t^i/b_i},
\;\;\;\;\;\;\;
E' = \prod_i \left[ \frac{\alpha_i(t)}{\alpha_i(X_2)}
\right]^{-c_t^i/b_i^{3\bar{3}}},
\;\;\;\;\;\;
E'' = \prod_i \left[ \frac{\alpha_i(t)}{\alpha_i(X_3)}
\right]^{-c_t^i/b_i^{MSSM}}.
\end{equation}

The expression of the wave-function renormalization may be easily obtained
by solving Eq. (\ref{wave}), making use of the above expressions. 
The derivatives of $\ln Z_Q$ with respect to $X_I$ may be obtained from the
formulae given in the appendix. For instance, 

\begin{eqnarray}
\frac{\partial \ln Z_Q}{\partial \ln X_2} &  = &
\frac{ c_t^i B_2^i d_t^Q \alpha_t(X_2) F'(X_3) H'(\alpha_i^2)}
{(4\pi)^3 \left[ 1 - \frac{ 6 \alpha_t(X_2) F'(X_3)}{4\pi}\right]}
+ 2 c_Q^i B_2^i \int_{X_2}^{X_3} \frac{\alpha_i^2}{\left(4\pi\right)^2} dt
\nonumber\\
& + & \frac{\alpha_t(X_3) F''(\mu)d_t^Q c_t^i B_2^i}
{(4\pi)^3 \left[ 1 - \frac{ 6 \alpha_t(X_3) F''(\mu)}{4\pi}\right]}
\left[ \int_{X_2}^{X_3} \alpha_i^2 + \frac{6 \alpha_t(X_2) F'(X_3) 
H'(\alpha_i^2)}
{4\pi \left(1 - \frac{ 6 \alpha_t(X_2) F'(X_3)}{4\pi}\right)} \right]
\nonumber\\
& + & \frac{ c_t^i B_2^i d_t^Q \alpha_t(X_3) F''(\mu) H''(\alpha_i^2)}
{(4\pi)^3 \left[ 1 - \frac{ 6 \alpha_t(X_3) F''(\mu)}{4\pi}\right]}
+ 2 c_Q^i B_2^i \int_{X_3}^{\mu} \frac{\alpha_i^2}{\left(4\pi\right)^2} dt.
\end{eqnarray}
The form of the linear integral functions 
$H'(f(t))$ and $H''(f(t))$ are given in the appendix.
In order to extract the trilinear mass parameter governing the
stop mixing, it is more useful to write
\begin{eqnarray}
\sum_{Q,U,H_2} \frac{\partial \ln Z_Q}{\partial \ln X_2} &  = &
\frac{c_t^i B_2^i}{(4\pi)^2} \left[ 
\frac{  6 \alpha_t(X_2) F'(X_3) H'(\alpha_i^2)}
{(4\pi) \left[ 1 - \frac{ 6 \alpha_t(X_2) F'(X_3)}{4\pi}\right]}
+ \int_{X_2}^{X_3} \alpha_i^2 \; dt\right]
\nonumber\\
& \times & \left[ 1 + \frac{6 \alpha_t(X_3) F''(\mu)}
{(4\pi) \left[ 1 - \frac{ 6 \alpha_t(X_3) F''(\mu)}{4\pi}\right]} \right]
\nonumber\\
& + & \frac{c_t^i B_2^i}{(4\pi)^2}  \left[
\frac{ 6 \alpha_t(X_3) F''(\mu) H''(\alpha_i^2)}
{(4\pi) \left[ 1 - \frac{ 6 \alpha_t(X_3) F''(\mu)}{4\pi}\right]}
+ \int_{X_3}^{\mu} \alpha_i^2 \; dt\right] .
\label{zqx2hqu}
\end{eqnarray}
Analogously,
\begin{equation}
\sum_{Q_L,U_R,H_2} \frac{\partial \ln Z_Q}{\partial \ln X_3}  = 
 \frac{c_t^i B_3^i}{(4\pi)^2}  \left[
\frac{ 6 \alpha_t(X_3) F''(\mu) H''(\alpha_i^2)}
{(4\pi) \left[ 1 - \frac{ 6 \alpha_t(X_3) F''(\mu)}{4\pi}\right]}
+ \int_{X_3}^{\mu} \alpha_i^2 \; dt \right] .
\label{zqx3hqu}
\end{equation} 

Let us rewrite the
formulae derived above in a  clearer way, by
introducing the quantities
\begin{eqnarray}
y' & =  & - \frac{ 6 \alpha_t(X_2) F'(X_3)}
{(4\pi) \left[ 1 + \frac{ 6 \alpha_t(X_2) F'(X_3)}{4\pi}\right]} ,
\nonumber\\
y'' & = & - \frac{ 6 \alpha_t(X_3) F'(\mu)}
{(4\pi) \left[ 1 + \frac{ 6 \alpha_t(X_3) F'(\mu)}{4\pi}\right]} .
\end{eqnarray}
The function $y'$  denotes the square of the ratio of the
Yukawa coupling at the scale $X_3$ to the value that it would
acquire at the scale $X_3$ if the coupling at $X_2$ were
strong. Similarly, $y''$ is the square of the
ratio of the Yukawa coupling at the scale $\mu$ to the value
that it would acquire at the scale $\mu$ if the coupling at
the scale $X_3$ were strong.
The functions $y'$ and $y''$ are analogous to the function $y$,
equal to the ratio of the top Yukawa coupling to its fixed point
value, which plays a relevant role in the solution of the renormalization
group equations in models in which supersymmetry breaking is
mediated by gravity effects \cite{IL},\cite{COPW},\cite{LT}.

The expression for the parameter $A_t = \sum_{Q=Q_L,U_R,H_2}
\partial \ln Z_Q/\ln X_I \times F_I/X_I$ becomes 
\begin{eqnarray}
A_t(\mu) & = & \frac{c_t^i M_i(X_2)}{4 \pi \alpha_i(X_2)}
\left[ \int_{X_2}^{X_3} \alpha_i^2(t) dt - y' H'(\alpha_i^2)
\right] \times \left( 1 - y'' \right)
\nonumber\\
& + &
\frac{c_t^i M_i(X_3) }{4 \pi \alpha_i(X_3)}
\left[ \int_{X_3}^{\mu} \alpha_i^2(t) dt - y'' H''(\alpha_i^2)
\right] .
\label{A_t}
\end{eqnarray}
The above expression has a clear interpretation: at the scales
$X_2$, $X_3$, the trilinear coupling takes values
\begin{eqnarray}
A_t(X_2) & = & 0 \;, 
\nonumber\\
A_t(X_3) & = & 
\frac{c_t^i M_i(X_2) }{4 \pi \alpha_i(X_2)}
\left[ \int_{X_2}^{X_3} \alpha_i^2(t) dt - y' H'(\alpha_i^2)
\right] .
\label{atx3}
\end{eqnarray}
There is no one-loop threshold contribution to the couplings
$A_t$ and the value at $X_3$ is  generated from the running
from the scales $X_2$ and $X_3$. The value of $A_t(X_3)$ then serves
as a boundary condition for the running between $X_3$ and
$\mu$ which generate both the $(-y'')$ factor in the first
line of Eq. (\ref{A_t}), as well as the whole second line.

Similar methods may be used to extract the soft supersymmetry 
breaking scalar mases from the second derivatives with respect
to the $X_I$'s of the 
chiral superfield wave-function renormalization. 
Keeping only the
terms depending on the top Yukawa coupling, we obtain
\begin{eqnarray}
\left.
-\frac{\partial \ln Z_Q}{\partial \ln X_2 \partial 
\ln X_2^{\dagger}} \right|_{\alpha_t} & = & \frac{d_Q^t}{4\pi} \left\{
 \frac{c_t^i\alpha_i(X_2)^2 B_2^i \alpha_t(X_2) F'(X_3) }
{(4\pi)^2 \left[ 1 - \frac{ 6 \alpha_t(X_2) F'(X_3)}{4\pi}\right]}
\right.
\nonumber\\
& + & \left.
\frac{2 \alpha_t(X_2) F'(X_3) c_t^j B_2^j c_t^i B_2^i  
H'\left(\alpha^2_j(t)\int_{X_2}^{\mu}
\alpha_i^2(t') dt'\right) }
{(4\pi)^4 \left[ 1 - \frac{ 6 \alpha_t(X_2) F'(X_3)}{4\pi}\right]}
\right.
\nonumber\\
& - &\frac{2 \alpha_t(X_2) F'(X_3) c_t^i (B_2^i)^2 H'(\alpha_i^3(t))}
{(4\pi)^3 \left[ 1 - \frac{ 6 \alpha_t(X_2) F'(X_3)}{4\pi}\right]}
\nonumber\\
& + & \left.
\frac{6 \alpha_t^2(X_2) \left(F'(X_3)\right)^2
\left[c_t^i B_2^i  H'(\alpha_i^2(t))\right]^2}
{(4\pi)^5 \left[ 1 - \frac{ 6 \alpha_t(X_2) F'(X_3)}{4\pi}\right]^2}
\right\}
\nonumber\\
& + & {\rm running \;  between \; scales} \;\; X_3 \;\; {\rm and} \;\; \mu ,
\label{zqx2}
\end{eqnarray}
where the last line represents the running effects between the scale
$X_3$ and $\mu$, which, because of their lengthy analytical expression,
we shall not write it down explicitly. Its form will become clear after
giving the expression for $m_Q^2(\mu)$. 

Assuming real gaugino mass
parameters, additional contributions to the scalar mass parameters
are obtained through,
\begin{eqnarray}
\left.
-\frac{\partial \ln Z_Q}{\partial \ln X_2 \partial 
\ln X_3^{\dagger}} + \left(X_2 \leftrightarrow X_3 \right)\right|_{\alpha_t}
& = & \frac{d_Q^t}{4\pi} \left\{ 
\frac{\alpha_t(X_3) F''(\mu) c_t^i B_2^i c_t^j B_3^j H''\left(\alpha_j^2(t)
\right)} 
{(4\pi)^3 \left[ 1 - \frac{ 6 \alpha_t(X_3) F''(\mu)}{4\pi}\right]^2}
\right.
\nonumber\\
& \times &\left[\int_{X_2}^{X_3} \frac{\alpha_i^2}{4\pi} dt +
\frac{6 \alpha_t(X_2) F'(X_3) H'\left(\alpha_i^2(t)\right)}
{(4\pi)^2 \left[ 1 - \frac{ 6 \alpha_t(X_2) F'(X_3)}{4\pi}\right]} \right]
\nonumber\\
& + & \frac{2 \alpha_t(X_3) F''(\mu) c_t^j B_2^j c_t^i B_3^i
H''\left(\alpha^2_j(t)\int_{X_3}^t
\alpha_i^2(t)\right) }
{(4\pi)^4 \left[ 1 - \frac{ 6 \alpha_t(X_3) F''(\mu)}{4\pi}\right]}
\nonumber\\
& - &\frac{2 \alpha_t(X_3) F''(\mu) c_t^i B_2^i B_3^i H''(\alpha_i^3(t))}
{(4\pi)^3 \left[ 1 - \frac{ 6 \alpha_t(X_3) F''(\mu)}{4\pi}\right]}
\nonumber\\
& + &  \left.
\frac{6 \alpha_t^2(X_3) \left(F''(\mu)\right)^2
\left[c_t^i B_2^i  H''(\alpha_i^2(t))\right]
\left[c_t^j B_3^j  H''(\alpha_j^2(t))\right]}
{(4\pi)^5 \left[ 1 - \frac{ 6 \alpha_t(X_3) F''(X_3)}{4\pi}\right]^2}
\right\} 
\nonumber\\
& + & \left(X_2 \leftrightarrow X_3\right) .
\label{zqx2x3}
\end{eqnarray}
Finally, the expression of $\partial \ln Z_Q/\ln X_3 \ln X_3^{\dagger}$
can be obtained by changing in an obvious way
the relevant scales and $\beta_i$ function coefficients 
in the expression given between curly brackets in 
Eq. (\ref{zqx2}).
The mass parameter $m_Q^2(\mu)$  can be obtained by
using the above expressions. At the scale $X_3$, for instance,
taking into account Eqs.~(\ref{mq2}),   
(\ref{mq2xi}) and (\ref{zqx2}), one obtains
\begin{eqnarray}
m_Q^2(X_3) & = & m_Q^2(X_2) 
- 2 c_Q^i \frac{M_i^2(X_2)}{\alpha_i^2(X_2)} 
\left[\frac{\alpha_i^2(X_3) - \alpha_i^2(X_2)}{b_i^{3\bar{3}}}
\right]
\nonumber\\
& - &\frac{y' d_Q^t}{6} \left(m_{Q_L}^2(X_2) + m_{U_R}^2(X_2)
+ m_{H_2}^2(X_2) \right)
\nonumber\\ 
& - & \frac{2 d_Q^t y'}{6} \frac{c_t^j M_j(X_2) c_t^i M_i(X_2)}
{\alpha_i(X_2) \alpha_j(X_2)} H'\left( \frac{\alpha_j^2}{(4 \pi)^2} 
\int_{X_2}^{\mu}
\alpha_i^2(t') dt' \right)
\nonumber\\
& + & \frac{2 d_Q^t y'}{6} \frac{c_t^i M_i^2(X_2)}{\alpha_i^2(X_2)}
H'\left(\frac{\alpha_i^3}{4\pi}\right)
+ \frac{d_Q^t}{6} \left(\frac{y' c_t^i M_i(X_2)}{\alpha_i(X_2)}
H'\left(\frac{\alpha_i^2}{4\pi}\right) \right)^2
\nonumber\\
& + & \frac{2 c_Q^i}{16 \pi^2} B_3^i \alpha_i^2(X_3)
\left( \frac{F_3}{X_3} \right)^2 ,
\end{eqnarray}
where the last term is, as said before, the threshold contribution
at the scale $X_3$ and the other terms, apart from $m_Q^2(X_2)$, 
represent the running between
the scales $X_2$ and $X_3$.

Now, inspection of the result for 
$\partial^2 \ln Z_Q/\ln X_2 \ln X_3^{\dagger}$
shows that there is now 
a new contribution absent in the running between $X_2$ and
$X_3$. This is  due to the appearence of a non-vanishing value
of the trilinear soft supersymmetry-breaking parameter at the
scale $X_3$,
$A_t(X_3) \neq 0$ (see Eq. (\ref{atx3})). Actually,
\begin{eqnarray}
m_Q^2(\mu) & = & m_Q^2(X_3) 
- 2 c_Q^i \frac{M_i^2(X_3)}{\alpha_i^2(X_3)} 
\left[\frac{\alpha_i^2(\mu) - \alpha_i^2(X_3)}{b_i^{MSSM}}
\right]
\nonumber\\
& - &\frac{y'' d_Q^t}{6} \left(m_{Q_L}^2(X_3) + m_{U_R}^2(X_3)
+ m_{H_2}^2(X_3) \right)
\nonumber\\ 
& - & \frac{2 d_Q^t y''}{6} \frac{c_t^j M_j(X_3) c_t^i M_i(X_3)}
{\alpha_i(X_3) \alpha_j(X_3)} H''\left( \frac{\alpha_j^2}{(4\pi)^2} 
\int_{X_3}^{\mu}
\alpha_i^2(t') dt' \right)
\nonumber\\
& + & \frac{2 d_Q^t y''}{6} \frac{c_t^i M_i^2(X_3)}{\alpha_i^2(X_3)}
H''\left(\frac{\alpha_i^3}{4\pi}\right)
+ \frac{d_Q^t}{6} \left(\frac{y'' c_t^i M_i(X_3)}{\alpha_i(X_3)}
H''\left(\frac{\alpha_i^2}{4\pi}\right) \right)^2
\nonumber\\
& - &  \frac{d_Q^t y'' ( 1- y'')}{6} \frac{2 c_t^i M_i(X_3)}{\alpha_i(X_3)}
H''\left(\frac{\alpha_i^2}{4\pi}\right) A_t(X_3)  
\nonumber\\
& - & \frac{d_Q^t y'' (1 - y'')}{6} A_t(X_3)^2 .
\label{scalarX3mu}
\end{eqnarray}
Observe that the last term in Eq. (\ref{scalarX3mu}) appears only through
the contribution to $\partial \ln Z_Q/\partial \ln X_2 \partial 
\ln X_2^{\dagger}$ omitted in Eq. (\ref{zqx2}). In order to understand
the origin of this term it is easier to study the 
contributions to
the soft supersymmetry breaking parameter $\bar{m}^2 = m_{Q_L}^2 +
m_{H_2}^2 + m_{U_R}^2$.  
It is easy to see
that the
expression given between curly brackets 
in Eq. (\ref{zqx2}) (suming over $Q = H_2,U_R$ and $Q_L$)
comes from the derivative of
the first line of Eq. (\ref{zqx2hqu})
with respect to $X_2^{\dagger}$. The derivative with 
respect to $X_2^{\dagger}$ of the last line of  Eq. (\ref{zqx2hqu})
may be obtained by using
Eq. (\ref{zqx2x3}) and observing that the last line in Eq. (\ref{zqx2hqu})
differs from Eq. (\ref{zqx3hqu}) only in the replacement of $B_3^i$ by
$B_2^i$. Finally, the contribution from the second line may be easily
obtained by using the following relation
\begin{eqnarray}
\frac{\partial y''}{\partial \ln X_2} &  = &- y'' (1 - y'') \times \left[ 
\frac{c_t^i B_2^i H''(\alpha_i^2)}{(4 \pi)^2}
\right.
\nonumber\\ 
& + &
\left.
\frac{c_t^i B_2^i }{(4\pi)^2} \left( \int_{X_2}^{X_3} \alpha_i^2 dt
- y' H' (\alpha_i^2) \right) \right].
\label{at2}
\end{eqnarray}
Using Eq. (\ref{atx3}) and the fact that the top Yukawa
contributions to $m_Q^2$ are weighted by the factors $d_Q^t$, the 
appearence of the 
last term in Eq. (\ref{scalarX3mu}) becomes clear.

\section{General Formulae}

After the above example, we can give general formulae for
the extraction of soft supersymmetry-breaking parameters in
the case of split messengers.

In general, in a theory with $\bar{N}$ thresholds 
$y^{(n)}$ can be defined as the
ratio of the square of the Yukawas at one threshold with
respect to the value that  would be obtained if the Yukawa coupling at
the immediately upper threshold were large,
\begin{equation}
y^{(n)} = - \frac{ 6 \alpha_t(X_{N}) F^{(n)}(X_{(N+1)})}
{(4\pi) \left[ 1 - \frac{ 6 \alpha_t(X_N) F^{(n)}(X_{(N+1)})}{4\pi}\right]} ,
\end{equation}
where $X_{N}$ 
is the n-th mass threshold scale and
$X_{N} > X_{(N+1)}$.
The function $F^{(n)}$ is defined by
\begin{equation}
F^{(n)}(X_{(N+1)})
= \int_{X_{N}}^{X_{(N+1)}} E^{(n)}(t) dt, \;\;\;\;\;\;\;\;\;
E^{(n)}(t) = \prod_i \left( \frac{\alpha_i(t)}{\alpha_i(X_{N})}
\right)^{-c_t^i/b^n_i} ,
\end{equation}
where $b^n_i$ is the $\beta_i$ function coefficient 
of the effective theory defined in the energy
range between the scales $X_{N}$ and $X_{(N+1)}$.
We shall also define 
the linear integral function 
$H^{(n)}$  by
\begin{equation}
H^{(n)}(f) = \int_{X_{N}}^{X_{(N+1)}} f(t) dt - 
\frac{1}{F^{(n)}(X_{(N+1)})}
\int_{X_{N}}^{X_{(N+1)}} F^{(n)}(t) f(t) dt,
\end{equation}
which is most useful to express the soft supersymmetry-breaking
parameters at the weak scale in a compact form.

Generalizing the  formulae presented in the last
section, the value of the gaugino masses
at the threshold $X_I$ is given by,
\begin{eqnarray}
M_j(X_1)  & = &  \frac{B_{1}^j\alpha_j(X_{1})}{4\pi}  
\frac{F_{1}}{X_{1}}
\nonumber\\
M_j(X_{(N+1)}) & = & M_j(X_{N}) \frac{\alpha_j(X_{(N+1)})}
{\alpha_j(X_{N})} +
           \frac{B_{(N+1)}^j\alpha_j(X_{(N+1)})}{4\pi}  
\frac{F_{(N+1)}}{X_{(N+1)}}
\nonumber\\
M_j(\mu) & = & M_j(X_{\bar{N}}) \frac{\alpha_j(\mu)}
{\alpha_j(X_{\bar{N}})},
\end{eqnarray}
where $B_N^i$ represents the contributions to the coefficient of the
function $\beta_i$ 
of the messenger with mass $X_N$.
As in the above examples, the gaugino masses receive
threshold contributions at each messenger mass scale and
their simple renormalization group evolution makes the
general expression for their values at low energies
rather simple. In particular,
the gaugino masses are not affected by the Yukawa coupling 
of the top quark at this order.

The trilinear soft supersymmetry-breaking mass terms, instead,
not only depend on the gauge sector, but also are affected
by top quark Yukawa dependent effects, that modify  their
renormalization group evolution. Following the example 
given in the last section we find for the general case, 
at this order:
\begin{eqnarray}
A_t(X_1) & = & 0
\nonumber\\
A_t(X_{(N+1)}) & = & 
\frac{c_t^i M_i(X_{N}) }{4 \pi \alpha_i(X_{N})}
\left[ \int_{X_N}^{X_{(N+1)}} \alpha_i^2(t) dt - y^{(n)} H^{(n)}(\alpha_i^2)
\right]
\nonumber\\
& + & A_t(X_{N}) \left( 1 - y^{(n)} \right) \; . 
\label{atev}
\end{eqnarray}
Since there are no threshold contributions to the trilinear soft
supersymmetry-breaking parameters,
the relation between the parameters 
$A_t(X_{\bar{N}})$ and $A_t(\mu)$ is governed by the same expression 
as governs the evolution between two different threshold scales,
Eq.~(\ref{atev}). 

The scalar mass parameters are also affected by gauge and 
top-quark Yukawa
contributions. The Yukawa dependence enters only through the
renormalization group evolution of these parameters. 
In the case of real gaugino masses, the general
expression is given by
\begin{eqnarray}
m_Q^2(X_1) & = &  \frac{2 c_Q^i}{16 \pi^2} B_{1}^i 
\alpha_i^2(X_{1})   \left( \frac{F_{1}}{X_{1}} \right)^2 , 
\nonumber\\
m_Q^2(X_{(N+1)}) & = & m_Q^2(X_{N}) 
- 4 c_Q^i \frac{M_i^2(X_{N})}{\alpha_i^2(X_{N})} 
\times \int_{X_{N}}^{X_{(N+1)}} \frac{\alpha_i^3(t)}{4\pi} dt
\nonumber\\
& - &\frac{y^{(n)} d_Q^t}{6} \left(m_{Q_L}^2(X_{N}) + m_{U_R}^2(X_{N})
+ m_{H_2}^2(X_{N}) \right)
\nonumber\\ 
& - & \frac{2 d_Q^t y^{(n)}}{6} \frac{c_t^j M_j(X_{N}) 
c_t^i M_i(X_{N})}
{\alpha_i(X_{N}) \alpha_j(X_{N})} 
H^{(n)}\left( \frac{\alpha_j^2(t)}{(4\pi)^2} \int_{X_{N}}^{\mu}
\alpha_i^2(t') dt' \right)
\nonumber\\
& + & \frac{2 d_Q^t y^{(n)}}{6} 
\frac{c_t^i M_i^2(X_{N})}{\alpha_i^2(X_{N})}
H^{(n)}\left(\frac{\alpha_i^3}{4\pi}\right)
+ \frac{d_Q^t}{6} \left(\frac{y^{(n)} 
c_t^i M_i(X_{N})}{\alpha_i(X_{N})}
H^{(n)}\left(\frac{\alpha_i^2}{4\pi}\right) \right)^2
\nonumber\\
& - & \frac{d_Q^t y^{(n)} ( 1- y^{(n)})}{6} 
\frac{2 c_t^i M_i(X_{N})}{\alpha_i(X_{N})}
H^{(n)}\left(\frac{\alpha_i^2}{4\pi}\right) A_t(X_{N})
\nonumber\\
& - & \frac{d_Q^t y^{(n)} ( 1 - y^{(n)})}{6} A_t(X_N)^2
\nonumber\\
& + & \frac{2 c_Q^i}{16 \pi^2} B_{(N+1)}^i \alpha_i^2(X_{(N+1)})  
\left( \frac{F_{(N+1)}}{X_{(N+1)}} \right)^2  ,
\end{eqnarray}
while the relation between $m_Q^2(X_{\bar{N}})$ and $m_Q^2(\mu)$ is analogous
to the one between $m_Q^2(X_3)$ and $m_Q^2(\mu)$ given in 
Eq. (\ref{scalarX3mu}), by changing $X_3$ by $X_{\bar{N}}$ and the 
corresponding $\beta_i$ functions in an obvious way.

Finally, let us write the expression for the $B_H$ parameter,
which governs the relation between the bilinear terms in the
superpotential and the bilinear soft supersymmetry-breaking
parameters in the scalar potential. This is given by
\begin{eqnarray}
B_H(X_1) & = & 0,
\\
B_H(X_{(N+1)}) & = & B_H(X_{N}) - \frac{y(n) d^t_{H_2}}{6}
A_t(X_N)
\nonumber\\
& + & \frac{4 c_{H_2}^i M_i(X_N)}{4 \pi \alpha_i(X_N)} 
\times \int_{X_N}^{X_{(N+1)}} \alpha_i^2(t) \; dt
 -\frac{ c_t^i M_i(X_N)}{4 \pi \alpha_i(X_N)}
\frac{y^{(n)} d_{H_2}^t}{6} H^{(n)}(\alpha_i^2) ,
\nonumber
\end{eqnarray}
with an analogous relation holding for the evolution between
the scales $X_{\bar{N}}$ and $\mu$. In general, however, we expect 
$B_H(X_1)$ to be modified by the dynamics leading to the
supersymmetric mass parameter $\mu$ 
in the superpotential \cite{muproblem}, and
hence in phenomenological analysis, $B_H(X_1)$ (or $B_H$ in general)
may be taken as a free parameter to be fixed by the requirement
of consistent radiative electroweak symmetry breaking. Observe
that the same dynamics might modify the Higgs mass
parameters $m_{H_i}^2$. These corrections, if present, will change
the boundary conditions of the parameters $m_{H_i}^2$ at the relevant
scale, but will preserve the form of the evolution of the soft 
supersymmetry breaking parameters to low energies.

\section{Conclusions}

In this work, we presented  formulae for the soft 
supersymmetry-breaking parameters in the observable sector
in gauge-mediated supersymmetry-breaking scenarios, for the
case of arbitrary quantum numbers of the messenger superfields.
The expressions for the low-energy soft supersymmetry-breaking
parameters were obtained by using the recently proposed method
to extract supersymmetry-breaking parameters from wave-function
renormalization.
The full dependence on the gauge and top-quark Yukawa coupling
effects was presented.
These formulae are particularly
useful to study the radiative breaking of the electroweak symmetry
in the regime of low and moderate values of $\tan\beta$
\footnote{The Higgs soft supersymmetry-breaking parameters at the
messenger mass scales should be corrected by the finite
next-to-leading order contributions, in case they are sizeable.
Moreover, in order to cancel the scale and scheme dependence of
these parameters, the full one-loop effective potential must
be considered~\cite{RG}.}.

In the most general case, the gaugino masses are very different
from those that would be obtained starting from universal values
at the unification scale. The generalized expressions for the
low energy soft supersymmetry breaking parameters hence make  contact
with similar expressions, presented  recently in the literature,
for the evolution of the supersymmetry breaking parameters for
arbitrary boundary conditions 
for the soft supersymmetry breaking parameters at a 
given scale $\Lambda$~\cite{CCOPW}. 
In fact,  the expression for the soft supersymmetry-breaking parameters
at low energies may be interpreted as resulting from the 
successive running between different threshold scales, defined
by the masses of the fermion messengers, plus the matching 
contributions coming from the decoupling of each messenger
superfield. The advantage of the method used in this article
is that it provides both the matching and the running contributions
without the risk of double-counting, or the need of explicit
evaluation of the Feynman diagrams. The expressions presented
in this article may be generalized to include the next-to-leading
order effects or off-diagonal $F$-terms. We reserve these improvements
for further analysis.

\vskip 2 cm
\noindent {\bf Acknowledgements}
~\\
\noindent  I would like to thank the Aspen Center for Physics and the
Argonne National Laboratory, U.S.A., and the University of Buenos Aires,
Argentina, where part of this work has been done. This work was
motivated by a project with T. Han and S. Mrenna, to whom I am
grateful. I would also like to thank S. Mrenna for very useful 
discussions at the beginning of this project.


\newpage
~\\
{\bf\large  Appendix}
\vskip 0.5cm

To obtain the dependence of the wave-function renormalization
on the different mass scales, the following formulae
are very useful :
\begin{eqnarray}
\frac{\partial \alpha_i(X_2)}
{\partial \ln X_2} & = & \frac{b_i}{4\pi} \alpha_i^2(X_2) ,
\;\;\;\;\;\;\;\;\;
\frac{\partial \alpha_i(X_3)}
{\partial \ln X_2}  =  \frac{b_i-b_i^{3\bar{3}}}{4\pi}
 \alpha_i^2(X_3) ,
\;\;\;\;\;\;\;
\frac{\partial \alpha_i(\mu)}{\partial \ln X_2} 
 =  \frac{b_i - b_i^{3\bar{3}}}{4\pi}
\alpha_i^2(\mu) ,
\nonumber\\
\frac{\partial F(X_2)}{\partial \ln X_2} & = & E(X_2), \;\;\;\;\;\;\;\;\;
\frac{
\partial \alpha_t(X_2)}{\partial \ln X_2} = -\frac{\alpha_t(X_2)}{4\pi}
\left(c_t^i \alpha_i(X_2) - 6 \alpha_t(X_2)\right) ,
\nonumber\\
\frac{\partial E(X_2)}{\partial \ln X_2}& = & - 
\frac{c_t^i \alpha_i(X_2)}{4\pi}
E(X_2) , \;\;\;\;\;\;\;\;\;\;\; 
\nonumber\\
\frac{\partial F'(X_3)}{\partial \ln X_2} &= & -1 
+ \frac{c_t^i b_i \alpha_i(X_2)}{4 \pi b_i^{3\bar{3}}} F'(X_3)
- \frac{c_t^i (b_i - b_i^{3\bar{3}})}{4\pi b_i^{3\bar{3}}} H'_i(X_2,X_3) ,
\nonumber\\
\frac{\partial E'(X_3)}{\partial \ln X_2} & = &
-\frac{c_t^i}{4\pi b_i^{3\bar{3}}} \left[ (b_i - b_i^{3\bar{3}})
\alpha_i(X_3) - b_i \alpha_i(X_2) \right] E'(X_3) ,
\nonumber\\
\frac{\partial \alpha_t(X_3)}{\partial \ln X_2} & = &
\frac{c_t^i (\alpha_i(X_3)-\alpha_i(X_2))}
{4 \pi} \left( 1 - \frac{b_i}{b_i^{3\bar{3}}}
\right) \alpha_t(X_3)   
- \frac{6 \alpha_t(X_3) \alpha_t(X_2) c_t^i (1-b_i/b_i^{3\bar{3}})}
{4\pi \left[4 \pi + 6 \alpha_t(X_2) F'(X_3)\right]}
\nonumber\\
& \times &
\left[ \alpha_i(X_2) F'(X_3) - H'_i(X_2,X_3) \right] ,
\nonumber\\
\frac{\partial E''(\mu)}{\partial \ln X_2} &=& -
\frac{c_t^i}{4\pi b_i^{MSSM}} (b_i - b_i^{3\bar{3}})
 \left(\alpha_i(\mu) - \alpha_i(X_3)
\right) E''(\mu) ,
\nonumber\\
\frac{\partial F''(\mu)}{\partial \ln X_2} & = & -
\frac{c_t^i (b_i-b_i^{3\bar{3}})}{4\pi b_i^{MSSM}}
\left[H''_i(X_3) - \alpha_i(X_3) F''(\mu)\right] ,
\nonumber\\
\frac{\partial H'_i(X_3)}{\partial \ln X_2} & = &
-\alpha_i(X_2) + \frac{b_i-b_i^{3\bar{3}}}{4\pi} H'_{ii}(X_2,X_3)
 +\frac{c_t^j b_j \alpha_j(X_2)}{4 \pi b_j^{3\bar{3}}} H'_i(X_2,X_3)
\nonumber\\
& - & \frac{c_t^j (b_j - b_j^{3\bar{3}})}{4\pi b_j^{3\bar{3}}}
H'_{ij}(X_2,X_3) ,
\end{eqnarray} 
where
\begin{eqnarray}
H'_i & = & \int_{X_2}^{X_3} \alpha_i(t) E'(t) dt ,
\;\;\;\;\;\;\;\;\;\;
H'_{ij} = \int_{X_2}^{X_3} \alpha_i(t) \alpha_j(t) E'(t) dt .
\end{eqnarray}
It is convenient to reexpress the above equations as a function of the
linear integral function
\begin{equation}
H'(f) = \int_{X_2}^{X_3} f(t) dt - \frac{1}{F'(X_3)}
\int_{X_2}^{X_3} F'(t) f(t) dt .
\end{equation}
Hence,
\begin{equation}
\int_{X_2}^{X_3} f(t) E'(t) dt = F'(X_3) \left[ H'\left(df/dt\right) +
f(X_2) \right] .
\end{equation}
Similar equations are found for the derivatives of the different
functions with respect to $\ln X_3$, with the simple facts that
$\partial f(X_2)/\partial \ln X_3 = 0$, and
\begin{eqnarray}
\frac{\partial \alpha_i(X_3)}
{\partial \ln X_3} &  = &  \frac{b_i^{3\bar{3}}}{4\pi}
 \alpha_i^2(X_3) ,
\;\;\;\;\;\;\;
\frac{\partial \alpha_i(\mu)}{\partial \ln X_2} 
 =  \frac{b_i^{3\bar{3}}- b_i^{MSSM}}{4\pi}
\alpha_i^2(\mu) ,
\nonumber\\
\frac{\partial F(X_3)}{\partial \ln X_3} & = & E(X_3), \;\;\;\;\;\;\;\;\;
\frac{
\partial\alpha_t(X_3)}{\partial \ln X_3} = -\frac{\alpha_t(X_3)}{4\pi}
\left(c_t^i \alpha_i(X_3) - 6 \alpha_t(X_3)\right) ,
\nonumber\\
\frac{\partial E(X_3)}{\partial \ln X_3}& = &- \frac{c_t^i \alpha_i(X_3)}{4\pi}
E(X_3) , \;\;\;\;\;\;\;\;\;\;\; 
\nonumber\\
\frac{\partial F''(\mu)}{\partial \ln X_3} &= & -1 
+ \frac{c_t^i b_i^{3\bar{3}} \alpha_i(X_3)}{4 \pi b_i^{MSSM}} F''(\mu)
- \frac{c_t^i (b_i^{3\bar{3}}-b_i^{MSSM})}{4\pi b_i{MSSM}} H''_i(X_3,\mu) ,
\nonumber\\
\frac{\partial E''(\mu)}{\partial \ln X_3} & = &
- \frac{c_t^i}{4\pi b_i^{MSSM}} \left[ (b_i^{3\bar{3}}-b_i^{MSSM})
\alpha_i(\mu) - b_i \alpha_i(X_3) \right] E''(\mu) ,
\nonumber\\
\frac{\partial \alpha_t(\mu)}{\partial \ln X_3} & = &
\frac{c_t^i (\alpha_i(\mu)-\alpha_i(X_3))}
{4 \pi} \left( 1 - \frac{b_i^{3\bar{3}}}{b_i^{MSSM}}
\right) \alpha_t(\mu)   
- \frac{6 \alpha_t(\mu) \alpha_t(X_3) c_t^i (1-b_i^{3\bar{3}}/b_i^{MSSM})}
{4\pi \left[4 \pi + 6 \alpha_t(X_3) F''(\mu)\right]}
\nonumber\\
& \times &
\left[ \alpha_i(X_3) F''(\mu) - H''_i(X_3,\mu) \right] ,
\nonumber\\
\frac{\partial H''_i(\mu)}{\partial \ln X_3} & = &
-\alpha_i(X_3) + \frac{b_i^{3\bar{3}}-b_i^{MSSM}}{4\pi} H''_{ii}(X_3,\mu)
+\frac{c_t^j b_j^{3\bar{3}} \alpha_j(X_3)}{4 \pi b_j^{MSSM}} H''_i(X_3,\mu)
\nonumber\\
& - & \frac{c_t^j (b_j^{3\bar{3}}-b_j^{MSSM})}{4\pi b_j^{MSSM}}
H''_{ij}(X_3,\mu) ,
\end{eqnarray}
where, much as in the previous case, we have defined the 
functions
\begin{eqnarray}
H''_i(X_3,\mu) & = & \int_{X_3}^{\mu} \alpha_i(t) E''(t) dt \; ,
\;\;\;\;\;\;\;
H''_{ij}(X_3,\mu) = \int_{X_3}^{\mu} \alpha_{i}(t) \alpha_j(t) E''(t) dt .
\end{eqnarray}
Once again, the equations take a simplified form if they are expressed 
as a function of the integral function
\begin{equation}
H''(f(t)) = \int_{X_3}^{\mu} f(t) dt - \frac{1}{F''(\mu)} 
\int_{X_3}^{\mu} F''(t) f(t) dt .
\end{equation}

\end{document}